\documentclass{ws-p9-75x6-50}
\begin{document}

\title{Colliding black holes from a null point of view: the close limit}

\author{Sascha Husa, Jeffrey Winicour}

\address{Department of Physics and Astronomy \\
         University of Pittsburgh, Pittsburgh, PA 15260\\ and\\
         Max-Planck-Institut f\" ur Gravitationsphysik,
         Albert-Einstein-Institut \\
         14476 Golm, Germany}

\author{Manuela Campanelli}

\address{Max-Planck-Institut f\" ur Gravitationsphysik,
         Albert-Einstein-Institut \\
         14476 Golm, Germany}  

\author{Roberto Gomez, Yosef Zlochower}

\address{Department of Physics and Astronomy \\
         University of Pittsburgh, Pittsburgh, PA 15260}  


\maketitle

\abstracts
{We present a characteristic algorithm for computing the
 perturbations of a
 Schwarzschild spacetime by means of solving the Teukolsky equations.
 Our methods and results  are
 expected to have direct bearing on the study of binary black holes presently
 underway using a fully {\em nonlinear} characteristic code \cite{Gomez98a}.
}

Recent years have shown a vital synergism between numerical and perturbative
approaches in the context of Cauchy evolution. In this work we try to make
these benefits also accessible in the context of null-surface-based evolution.
We treat the ``close-limit'' approximation to the post-merger phase of a
binary black hole spacetime in terms of the characteristic initial value
problem. 
We outline a framework to treat the Teukolsky equation ~\cite{Teukolsky73}
in a null evolution setting, and our numerical strategy
to solve the perturbative problem in the time domain. 
A detailed discussion and results for the outgoing radiation problem
are given in \cite{PaperI}, the details of the ingoing radiation problem
will be treated in a forthcoming paper \cite{PaperII}.

We apply our method to evolve ``close-limit''
horizon data, which we obtain from linearizing the data discussed in
\cite{Nulltube-paper}. There we use Bondi-Saschs coordinates, which
simplify to Israel coordinates in the spherical case, which are based on
an affine parameter $u$ along the white hole horizon, normalized
to $u=0$ at the bifurcation sphere, and an affine parameter $\lambda$
along the outgoing null cones, fixed to $\lambda=0$ at the white hole horizon
and $(\partial^a u) (\partial_a \lambda)=-1)$.
These coordinates cover the entire Kruskal manifold $r>0$ with remarkably
simple analytic behavior, as first discovered by Israel~\cite{israel}.
Using these convenient coordinates as a starting point, it is straightforward
to write the Teukolsky equations explicitly in any background coordinate
system.

In order to
treat the radiation near ${\cal I}^+$ it is  advantageous to consider a
``boosted tetrad''  $({\tilde l}^a, {\tilde n}^a, m^a, \bar m^a )$
with ${\tilde l}^a=-\nabla^a \tilde{u}$,  ${\tilde n}^a$
satisfying ${\tilde l}^a {\tilde n}_a = -1$ and a dyad
$ m^a=\frac{1}{{\sqrt 2} r}  (\frac{\partial}{\partial\vartheta})^a +
        \, \frac{i}{\sin\vartheta}(\frac{\partial}{\partial\varphi})^a$.
 We accordingly define boosted Weyl scalars ${\tilde \psi}_0 =
C_{abcd} {\tilde l}^a m^b {\tilde l}^c m^d$ and
 ${\tilde \psi}_4 = C_{abcd}{\tilde n}^a \bar m^b {\tilde n}^c \bar m^d$.
 This boosted tetrad is adapted to
the affine time on $\tilde u$ at ${\cal I}^+$ rather than the affine time $u$
at the horizon.

This tetrad has proven useful for the evolution of
outgoing radiation described by ${\tilde \psi}_4$, for ingoing radiation
described by ${\tilde \psi}_0$, a more sophisticated choice is necessary
to achieve optimal results, as will be discussed in \cite{PaperII}.

In order to accurately track the evolution of  ${\tilde \psi}_4$ well into
the power law tail phase (which typically requires more than a 1000 M)
of evolution in Bondi time $\tilde u$, we factor out the asymptotic falloff
of ${\tilde \psi}_4$ for large $r$, we use fixed time steps in Bondi time and
a compactified version $\rho$ of the tortoise coordinate
$r*=r + 2 M \ln(r/2M - 1)$, $\rho=\rho_0 \tan \rho$,
where $\rho_0$ is an adjustable parameter. The coordiante $\rho$ allows for
a fixed grid to accurately resolve the Schwarzschild potential at all times,
which proved vital in the tail phase.
For a typical choice of initial data  the power law  tail only sets in after
the quasinormal oscillations have decayed by more than 10 orders of magnitude.
In order for the final tail not to be lost in machine error it is necessary to
evolve the quasinormal phase in quadruple precision.

We have obtained preliminary results to construct close approximation
black hole null data without ingoing radiation, by using superposition of
solutions for the linear Teukolsky equation, and by utilizing the time
symmetry of the background Schwarzschild spacetime which implies a
correspondence map of a retarded solution (no incoming radiation) for $\Psi_4$
into an advanced solution (no outgoing radiation) for $\Psi_0$. Details
will appear in \cite{PaperII}.

\section*{Acknowledgments}
M.C. was partially supported by a Marie-Curie Fellowship (HPMF-CT-1999-00334)
This work has been partially supported by NSF PHY
9988663 and NSF PHY 9800731 to the University of
Pittsburgh. Computer time for this project was provided by the
Pittsburgh Supercomputing Center and by NPACI.


\end{document}